\documentclass[12pt,preprint]{aastex}

\begin{document}

\title{Accurate estimations of circumstellar and interstellar lines of quadruply ionized vanadium using the coupled cluster approach}

\author{Gopal Dixit$^1$, B. K. Sahoo$^2$, P.C. Deshmukh $^1$, R. K. Chaudhuri$^3$, Sonjoy Majumder $^1$} 
\affil{\it $^1$Department of Physics, Indian Institute of Technology-Madras, Chennai-600 036, India \\
$^2$ Max Planck Institute for the Physics of Complex Systems, D-01187 Dresden, 
Germany\\
$^3$ Indian Institute of Astrophysics, Bangalore-34, India}


\begin{abstract}
\noindent
Accurate {\it ab initio} calculations have been carried out to study the 
valence electron removal energies and oscillator strengths of astrophysically 
important electromagnetic transitions of quadruply ionized vanadium, $V^{4+}$. Many important 
electron correlations are considered to all-orders using the relativistic coupled-cluster theory. 
Calculated ionization potentials and fine structure splittings are compared with the experimental values, 
wherever available. To our knowledge, oscillator strengths of electric dipole transitions 
are predicted for the first time for most of the transitions. The transitions 
span in the range of ultraviolet, visible and near infrared regions and are important for
astrophysical observations. 
\end{abstract}


\section{Introduction}
Vanadium (V) is very important constituent in the atmosphere of dwarf stars \citep{kirk} as it condenses 
into solid solution with Ti-bearing condensates (as observed in meteorites) and not as 
pure vanadium oxide ($V_2O_5$) as commonly assumed. Abundance of V in its ionic
or molecular form is used to constrain temperature \citep{lodders} at the dwarf transition.
The abundance of V has been seen for 46 normal stars of spectral types from G8 to M0, 
and their accurate estimates depend on the precise evaluation of transition amplitudes. 
This abundance is very important in Sun \citep{bonsack} and the observed maximum abundance 
among similar stars declines with surface temperature. The abundance excesses of this odd Z 
element by 1.2 dex  in the hot peculiar star, like `star 3 cen A', has very interesting and important 
applications in astrophysics \citep{cowley}.  
The transition between low-lying energy levels is also of importance to the identification of 
V emission lines in solar spectra. Spectroscopic studies of this multi-charged ion are important
in numerous applications in laser and plasma physics apart from astrophysics, like, the operations of 
short-wave laser plasma, plasma diagnostics etc \citep{russian}.

The dominant configuration to the ground state of quadruply ionized vanadium ($V^{4+}$) is 
[Ar]$3d_{3/2}$ $\bigl(^2D_{3/2}\bigr)$, with strong correlations from $3d$ orbitals. 
This is unlike the isoelectronic neutral 
potassium (K), whose ground state is dominated by [Ar]$4s_{1/2}$ $\bigl(^2S_{1/2}\bigr)$. 
Experimental ionization potentials (IP) of ground and excited states of 
this ion \citep{deurzen} are tabulated in the NIST web database \citep{nist}. 
The comparison highlights the effective considerations of quantum 
many-body theories, especially the correlation contributions applied to evaluate these energy 
eigenstates. There is no line strength estimation among different low lying states obtained 
from the change of outer most valence orbitals of $V^{4+}$ to our knowledge. 
For few transitions, preliminary semi-empirical results are available wherein experimental energies had 
been used to calculate oscillator strengths \citep{russian1}.

Here we have employed the relativistic coupled cluster method with single and double excitations (RCCSD).
This is one of the most powerful highly correlated approaches due to its all order behavior for the 
correlation
operator \citep{lindgren}. The all order behavior comes from the generation of correlated
RCCSD states  using exponential structure of the cluster operators over the Dirac-Fock (DF) reference states 
\citep {coester}, which is explicitly explained in our recent paper \citep{sahoo2}. 
The contributions due to Breit interaction \citep{breit} which is four orders 
smaller than the Coulomb interaction have been neglected.
For a univalent atomic system a set of coupled equations for the cluster 
operators, $T_n$, may be obtained from the Bloch equations for core and valence electrons separately
\citep{lindgren,debasish}. 
Because of this consideration,
though the cluster operators are only of single and double excitations kind, their combinations can produce 
partial triple excitations which nominate the method as coupled cluster method with single, double and 
partial triple excitations (CCSD(T)).  Contributions from these partial excitations in the ionization 
potentials (IPs) for other systems are well known \citep{bartlett}; this technique has been described 
extensively in one of our recent papers \citep{sahoo2}.  Therefore, it can take into account of different electron 
correlations, like core correlation, core polarization and pair correlation,  effects exhaustively for 
specific type of excitations \citep{sahoo2}. 

The oscillator strength for a transition from $|\Psi_i\rangle$ to $|\Psi_f\rangle$ is
\begin{equation}
f_{fi} ={2\over {3g_i}}\Delta E_{fi}\times |D_{fi}|^2
\end{equation}
where $\Delta E_{fi}$ is the energy difference between final and initial states
 and $g_i=2J_i+1$ is the degeneracy factor of the initial state with angular momentum $J_i$. The electric dipole moment
matrix element $D_{fi}$ is defined as
\begin{equation}
D_{fi}\ =\ \frac{\langle \Psi_f|{\bf d}|\Psi_i\rangle}{\sqrt{\langle \Psi_f|\Psi_f\rangle}
\sqrt{\langle \Psi_i|\Psi_i\rangle}} \\
\end{equation}
where,
\begin{equation}
\langle \Psi_f|{\bf d}|\Psi_i\rangle\ =\ \langle\Phi_f|e^{T_f^{\dag}} de^{T_i}
|\Phi_i\rangle
\end{equation}
{\bf d}$=e \vec \textbf{r}$ is the electric dipole moment operator in the length ($L$) gauge form.
$|\Phi\rangle$s' are the DF reference states.
Similarly, transition matrix elements can be obtained by replacing {\bf d} in velocity gauge form \citep{lin,alex}.
The connected parts of this expression will contribute and hence we only compute
those parts in our dipole matrix element calculations.

In the astrophysical literature, one uses weighted oscillator strength which is the product of degeneracy of the initial state and the oscillator strength and is symmetric with respect to initial and final states. i.e.,
\begin{equation}
 gf = (2J_i+1)f_{if} = - (2J_f+1)f_{fi}.
\end{equation}

\section{Results and Discussion}

We have started with the generation of DF orbitals constructed as linear combination of Gaussian type 
orbitals (GTOs), given by Chaudhuri et al. \citep{rajat}.  The number of single particle orbital bases are chosen for 
different symmetries based on the number of occupied orbitals and requirement of the transitions
of interest. All the core orbitals are considered active in our calculations to make 
correlations more exhaustive, especially the core correlation part.  
Table I shows the excitation energies and fine structure splittings of  many low
lying states with single unfilled orbitals of $V^{4+}$. For ionization potentials, we find very good 
agreement with 
the experimental data obtained from  NIST database \citep{nist} for all the cases. 
Leaving out the $3d_{5/2}$ state which has rather low binding energy, the average deviation 
from the experimental values are about 0.1\%. The experimental uncertainty of the allowed transition
wavelengths in the NIST values are very small, at the level of 2nd to 3rd decimal places, which are
of the order 10 cm$^{-1}$ in excitation energies. The fine-structure splittings are of the same
order. Therefore, it is important to do higher precession experiments for fine-structure splittings. 
  
\begin{table}
\caption{Ionization Potentials(IPs) and fine structure (FS) splittings
(in cm$^{-1}$) of $V^{+4}$ and their comparison with NIST values. The percentage
of differences of our calculated IP results compared to NIST results are shown in 
parenthesis on the side of CCSD(T) IP results. }
\begin{tabular}{lrrrr}
             & \multicolumn{2}{c}{IP} & \multicolumn{2}{c}{FS} \\ \hline
States       &  CCSD(T)        & NIST       &  CCSD(T)  &  NIST  \\
             &                 &            &           &         \\
\hline
$3d_{3/2}$   &     000.00(0.00)    &    000.00  &           &         \\
$3d_{5/2}$   &     707.63(13.24)   &    624.87  &  707.63   & 624.87   \\
$4s_{1/2}$   &  147728.21(0.28)    & 148143.35  &           &          \\
$4p_{1/2}$   &  206489.02(0.04)    & 206393.72  &           &          \\
$4p_{3/2}$   &  207779.93(0.05)    & 207660.00  & 1290.91   & 1266.28  \\
$4d_{3/2}$   &  293925.55(0.00)    & 293902.86  &           &          \\
$4d_{5/2}$   &  294082.10(0.01)    & 294047.24  &  156.54   &  144.38  \\
$5s_{1/2}$   &  328103.19(0.03)    & 328217.30  &           &          \\
$5p_{1/2}$   &  351832.99(0.09)    & 351500.51  &           &          \\
$5p_{3/2}$   &  352318.34(0.08)    & 352018.34  &  485.34   &  517.83  \\
$5d_{3/2}$   &  388227.86(0.06)    & 387977.07  &           &          \\
$5d_{5/2}$   &  388300.23(0.06)    & 388043.69  &  72.37    &  66.62   \\ 
$6s_{1/2}$   &  404449.78(0.14)    & 403855.12  &           &          \\
$6p_{1/2}$   &  415612.86(0.04)    & 415420.10  &           &          \\
$6p_{3/2}$   &  415591.43(0.02)    & 415675.69  &   21.43   &   225.59 \\
$4f_{5/2}$   &  347524.11(0.61)    & 349675.57  &           &          \\
$4f_{7/2}$   &  347500.44(0.50)    & 349252.40  &  23.67    &   423.17 \\
$5g_{7/2}$   &  414969.56(0.33)     & 416360.29  &          &          \\
$5g_{9/2}$   &  414963.83(0.33)     & 416361.78  & 05.73    & 01.49   \\
$6d_{3/2}$   &  434623.00(0.07)    & 434303.77  &           &          \\
$6d_{5/2}$   &  434653.58(0.07)    & 434340.92  &   30.58   &  37.15   \\
$6g_{7/2}$   &  450047.94(0.00)    & 450024.54  &           &         \\
$6g_{9/2}$   &  450043.64(0.00)    & 450025.20  & 04.30     &  00.64   \\
\end{tabular}
\label{tab:results1}
\end{table}

The oscillator strength for transitions between different energy states depend linearly on  energy, and
quadratically on the electric dipole matrix elements (Eq. (1)). This shows the necessity to calculate 
transition matrix elements appropriately. We present the electric dipole transition amplitudes in both 
length and velocity gauges in Table II. The good agreement between these two forms from the calculations
using the same initial and final wavefunctions  highlights the fact that the numerical algorithm is robust. 
A detailed derivation and interpretation of different gauges can be found in \citep{bethe}, and in 
few recent references \citep{alex}. The little differences in the values of these two forms for particular
transition amplitude may be improved by considering negative energy states in the calculation. 
In few transition 
amplitudes, there are considerable disagreement between the length- and velocity-forms. The detail analysis
might be possible using the approach suggested by \citep{savukov}.
In table III, we have given the {\it ab initio} oscillator strengths (f-value) for 
`ground to excited' and `excited to excited' transitions, which are also astrophysically important. 
Here we are presenting the oscillator 
strength in the length gauge due to its comparative fast convergence \citep{idress}.

\begin{table}[h]
\caption{Comparisons between length and velocity gauge transition amplitudes.}
\begin{tabular}{llrrcllrr}
\hline
Terms     &                      &  Length  & Velocity & & Terms      &            &  Length  & Velocity \\ \hline
$4s_{1/2}$&$\rightarrow 4p_{1/2}$&  1.74965 &  1.75033 & &  $4d_{3/2}$&$\rightarrow 4p_{1/2}$&  2.76747 &  2.84837 \\ 
          &$\rightarrow 5p_{1/2}$&  0.23476 &  0.21403 & &            &$\rightarrow 5p_{1/2}$&  2.18131 &  1.99049 \\
          &$\rightarrow 6p_{1/2}$&  0.05003 &  0.03953 & &            &$\rightarrow 6p_{1/2}$&  0.10890 &  0.07280 \\
          &$\rightarrow 4p_{3/2}$& -2.47677 & -2.36660 & &            &$\rightarrow 4p_{3/2}$&  1.24409 &  1.19310 \\
          &$\rightarrow 5p_{3/2}$& -0.31660 & -0.38246 & &            &$\rightarrow 5p_{3/2}$&  0.96719 &  0.96266 \\
          &$\rightarrow 6p_{3/2}$& -0.07101 & -0.07757 & &            &$\rightarrow 6p_{3/2}$&  0.05241 &  0.05954 \\
$3d_{3/2}$&$\rightarrow 4p_{1/2}$& -0.70241 & -0.48866 & &            &$\rightarrow 4f_{5/2}$& -4.44233 & -4.51143 \\
          &$\rightarrow 5p_{1/2}$& -0.19265 & -0.07370 & &  $4d_{5/2}$&$\rightarrow 4p_{3/2}$&  3.73203 &  3.53595 \\
          &$\rightarrow 6p_{1/2}$& -0.09081 & -0.04798 & &            &$\rightarrow 5p_{3/2}$&  2.90946 &  2.93882 \\
          &$\rightarrow 4p_{3/2}$& -0.31147 & -0.28369 & &            &$\rightarrow 6p_{3/2}$&  0.14753 &  0.17376 \\
          &$\rightarrow 5p_{3/2}$&  0.08668 &  0.08420 & &            &$\rightarrow 4f_{5/2}$&  1.18770 &  1.20082 \\
          &$\rightarrow 6p_{3/2}$&  0.02889 & 0.10971 & &            &$\rightarrow 4f_{7/2}$&  5.30192 &  5.38146 \\
          &$\rightarrow 4f_{5/2}$&  0.67548 &  0.79536 & &  $4f_{5/2}$&$\rightarrow 5g_{7/2}$& -5.63172 & -5.19379 \\
$3d_{5/2}$&$\rightarrow 4p_{3/2}$& -0.93952 & -0.95399 & &            &$\rightarrow 6g_{7/2}$& -1.84114 & -1.55890 \\
          &$\rightarrow 5p_{3/2}$&  0.26166 &  0.34994 & &  $4f_{7/2}$&$\rightarrow 5g_{7/2}$& -1.08111 & -0.99647 \\
          &$\rightarrow 6p_{3/2}$&  0.14228 &  0.10591 & &            &$\rightarrow 6g_{7/2}$& -0.35363 & -0.29909 \\
          &$\rightarrow 4f_{5/2}$& -0.17964 & -0.19236 & &            &$\rightarrow 5g_{9/2}$&  6.39624 &  5.89684 \\
          &$\rightarrow 4f_{7/2}$& -0.81273 & -0.97714 & &            &$\rightarrow 6g_{9/2}$&  2.09246 &  1.77063 \\
$5s_{1/2}$&$\rightarrow 4p_{1/2}$&  0.91525 &  0.86305 & &  $5d_{3/2}$&$\rightarrow 4p_{1/2}$& -0.05739 & -0.15377 \\
          &$\rightarrow 5p_{1/2}$&  3.45262 &  3.41492 & &            &$\rightarrow 5p_{1/2}$&  5.14639 &  5.08534 \\
          &$\rightarrow 6p_{1/2}$& -0.09268 & -0.08414 & &            &$\rightarrow 6p_{1/2}$& -1.07982 & -0.96877 \\
          &$\rightarrow 4p_{3/2}$& -1.31827 & -1.35452 & &            &$\rightarrow 4p_{3/2}$& -0.01914 & -0.02498 \\
          &$\rightarrow 5p_{3/2}$& -4.87908 & -4.72249 & &            &$\rightarrow 5p_{3/2}$&  2.31138 &  2.20311 \\
          &$\rightarrow 6p_{3/2}$&  0.13169 &  0.14329 & &            &$\rightarrow 6p_{3/2}$& -0.48293 & -0.46061 \\
$6s_{1/2}$&$\rightarrow 4p_{1/2}$&  0.27420 &  0.24626 & &            &$\rightarrow 4f_{5/2}$& -2.44997 & -2.39345 \\
          &$\rightarrow 5p_{1/2}$& -1.58072 & -1.67907 & &  $5d_{5/2}$&$\rightarrow 4p_{3/2}$& -0.06172 &  0.09229 \\
          &$\rightarrow 6p_{1/2}$&  1.58456 &  1.31189 & &            &$\rightarrow 5p_{3/2}$&  6.93100 &  6.58011 \\
          &$\rightarrow 4p_{3/2}$& -0.39158 & -0.43908 & &            &$\rightarrow 6p_{3/2}$& -1.44693 & -1.38397 \\
          &$\rightarrow 5p_{3/2}$&  2.41224 &  2.34950 & &            &$\rightarrow 4f_{5/2}$&  0.65354 &  0.64360 \\
          &$\rightarrow 6p_{3/2}$& -2.22873 & -1.82955 & &            &$\rightarrow 4f_{7/2}$&  2.91305 &  2.85422 \\
$6d_{3/2}$&$\rightarrow 4p_{1/2}$& -0.03437 & -0.04741 & &  $6d_{5/2}$&$\rightarrow 4p_{3/2}$&  0.05564 &  0.16646 \\
          &$\rightarrow 5p_{1/2}$&  0.36454 &  0.37956 & &            &$\rightarrow 5p_{3/2}$&  0.46061 &  0.26962 \\
          &$\rightarrow 6p_{1/2}$&  2.32712 &  1.96753 & &            &$\rightarrow 6p_{3/2}$&  3.10432 &  2.56486 \\
          &$\rightarrow 4p_{3/2}$&  0.01925 &  0.05307 & &            &$\rightarrow 4f_{5/2}$&  0.11103 &  0.11702 \\
          &$\rightarrow 5p_{3/2}$&  0.15162 &  0.09272 & &            &$\rightarrow 4f_{7/2}$&  0.49502 &  0.51231  \\
          &$\rightarrow 6p_{3/2}$&  1.03434 &  0.85524 & & & & & \\
          &$\rightarrow 4f_{5/2}$& -0.41493 & -0.42212 & & & & & \\
\end{tabular}
\end{table}

\begin{table}[h]
\caption{ Transition wavelengths and oscillator strengths of $V^{4+}$.}
\begin{tabular}{llrrr}
\hline
Terms     &     &  $\lambda_{NIST}$(\AA) & $\lambda_{CCSD}$(\AA)  & $gf$-value\\ \hline
$4s_{1/2}$&$\rightarrow 4p_{1/2}$& 1716.72     & 1701.81     & 0.5464 \\
          &$\rightarrow 4p_{3/2}$& 1680.20    & 1665.23     &  1.1189  \\
          &$\rightarrow 5p_{1/2}$& 491.74     & 489.94      &  2.9499  \\  
          &$\rightarrow 5p_{3/2}$& 490.49     & 448.78      &  6.2293  \\
          &$\rightarrow 6p_{1/2}$& 374.14     & 373.29      &  2.0368  \\  
          &$\rightarrow 6p_{3/2}$& 373.78     & 373.32      &  0.0041  \\
$3d_{3/2}$&$\rightarrow 4p_{1/2}$& 484.51      & 484.28     &  0.3094  \\
          &$\rightarrow 4p_{3/2}$& 481.55      & 481.27     &  0.0612  \\
          &$\rightarrow 5p_{1/2}$& 284.49      & 284.22     &  0.8901  \\ 
          &$\rightarrow 5p_{3/2}$& 284.07      & 283.83     &  0.0080  \\
          &$\rightarrow 6p_{1/2}$& 240.72      & 240.60     &  0.0104 \\
          &$\rightarrow 6p_{3/2}$& 240.57      & 240.62     &  0.0010  \\
          &$\rightarrow 4f_{5/2}$& 285.97      & 287.74        & 0.4816 \\
$3d_{5/2}$&$\rightarrow 4p_{3/2}$& 483.00      & 482.92     &  0.5552  \\
          &$\rightarrow 5p_{3/2}$& 284.58      & 284.40     &  0.0731  \\
          &$\rightarrow 6p_{3/2}$& 240.93      & 241.03     &  0.0255  \\
          &$\rightarrow 4f_{5/2}$& 286.49      &288.33        & 0.0339 \\
          &$\rightarrow 4f_{7/2}$& 286.83      &288.35        & 0.6958 \\
$5s_{1/2}$&$\rightarrow 4p_{1/2}$& 820.85      & 822.27     &  0.3094  \\
          &$\rightarrow 4p_{3/2}$& 829.48      & 831.09     &  0.6351  \\
          &$\rightarrow 5p_{1/2}$& 4294.94     & 4214.11    &  0.8592  \\
          &$\rightarrow 5p_{3/2}$& 4201.49     & 4129.64    &  1.7510  \\
          &$\rightarrow 6p_{1/2}$& 1146.75     & 1142.73    &  0.0022  \\
          &$\rightarrow 6p_{3/2}$& 1143.40     & 1143.01      &0.0040  \\
$6s_{1/2}$&$\rightarrow 4p_{1/2}$&  506.42     & 505.15       &0.0452\\
          &$\rightarrow 4p_{3/2}$& 509.69      & 508.46       &0.0916\\
          &$\rightarrow 5p_{1/2}$& 1910.05     & 1900.53      &0.3993\\
          &$\rightarrow 5p_{3/2}$& 1929.13     & 1918.22      &0.9214\\
          &$\rightarrow 6p_{1/2}$& 8646.79     & 8958.09      & 0.0851\\
         &$\rightarrow 6p_{3/2}$& 8459.82     & 8975.33      & 0.1681\\
$4f_{5/2}$&$\rightarrow 5g_{7/2}$& 1499.59     &1541.18       & 6.2510\\
          &$\rightarrow 6g_{7/2}$& 996.52      &959.71        & 1.0728\\
\hline
\label{tab:front3}
\end{tabular}
\end{table}
\begin{table}[h]
\begin{tabular}{llrrr}
\hline
Terms     &     &  $\lambda_{NIST}$(\AA) & $\lambda_{CCSD}$(\AA)  & $gf$-value\\ \hline
$4d_{3/2}$&$\rightarrow 4p_{1/2}$&  1142.73     & 1143.68      &2.0341\\
          &$\rightarrow 4p_{3/2}$& 1159.51     & 1160.82      &0.4050\\
          &$\rightarrow 5p_{1/2}$& 1736.18     & 1726.89      & 0.8369\\
          &$\rightarrow 5p_{3/2}$& 1720.71     &1712.54       & 0.1659\\
          &$\rightarrow 6p_{1/2}$& 822.92      & 821.77       & 0.0043\\
          &$\rightarrow 6p_{3/2}$& 821.53      & 821.92       & 0.0010\\
          &$\rightarrow 4f_{5/2}$& 1792.99     &1865.72       & 3.2129\\
$4d_{5/2}$&$\rightarrow 4p_{3/2}$& 1157.57      & 1158.71      &3.6512\\
          &$\rightarrow 5p_{3/2}$& 1724.99     &1717.14       & 1.4974\\
          &$\rightarrow 6p_{3/2}$& 822.17      &822.98        & 0.0080\\
          &$\rightarrow 4f_{5/2}$& 1797.64     &1871.18       &0.2289\\
          &$\rightarrow 4f_{7/2}$& 1811.42     &1872.01       & 4.5612\\
$4f_{7/2}$&$\rightarrow 5g_{7/2}$& 1490.13     &1540.61       & 0.2304\\
          &$\rightarrow 5g_{9/2}$& 1490.10     &1446.24       & 8.5927\\
          &$\rightarrow 6g_{7/2}$& 992.33      &959.49        & 0.0395\\
          &$\rightarrow 6g_{9/2}$& 992.33      &975.19        &1.3638 \\
$5d_{3/2}$&$\rightarrow 4p_{1/2}$& 550.71      &550.24        &0.0018\\
          &$\rightarrow 4p_{3/2}$& 554.57      &554.17        &0.0002\\
          &$\rightarrow 5p_{1/2}$& 2741.48     &2747.64       &2.9279\\
          &$\rightarrow 5p_{3/2}$& 2780.96     &2784.77       &0.5827\\
          &$\rightarrow 6p_{1/2}$& 3643.91     &3651.63       &0.0969\\
          &$\rightarrow  6p_{3/2}$& 3610.28     &3654.49       &0.0193\\
          &$\rightarrow 4f_{5/2}$& 2610.86     &2656.77       &0.7421\\
$5d_{5/2}$&$\rightarrow 4p_{3/2}$& 554.37      &553.95        &0.0020\\
          &$\rightarrow 5p_{3/2}$& 2775.82     &2779.17       &5.2505\\
          &$\rightarrow 6p_{3/2}$& 3618.99     &3664.18       &0.1735\\
          &$\rightarrow 4f_{5/2}$& 2606.33     &2452.41       &0.0529\\
          &$\rightarrow 4f_{7/2}$& 2577.89     &2450.98       &1.0516 \\
$6d_{3/2}$&$\rightarrow 4p_{1/2}$& 438.76      &438.33        &0.0008\\
          &$\rightarrow 4p_{3/2}$& 441.22      &440.83        &0.0002\\
          &$\rightarrow 5p_{1/2}$& 1207.68     &1207.87       &0.0334\\
          &$\rightarrow 5p_{3/2}$& 1215.28     &1214.99       &0.0057\\
          &$\rightarrow 6p_{1/2}$& 5295.58     &5260.35       &0.3127\\
          &$\rightarrow 6p_{3/2}$& 5368.23     &5254.42       &0.0618\\
          &$\rightarrow 4f_{5/2}$& 1181.63     &1148.12       &0.0455\\
$6d_{5/2}$&$\rightarrow 4p_{3/2}$& 441.14      &440.83        &0.0021\\
          &$\rightarrow 5p_{3/2}$& 1214.73     &1214.98       &0.0530\\
          &$\rightarrow 6p_{3/2}$& 5360.42     &5254.26       &0.5571\\
          &$\rightarrow 4f_{5/2}$& 1181.26     &1148.11       &0.0032\\
          &$\rightarrow 4f_{7/2}$& 1175.38     &1147.80       &0.0648 \\
\hline
\label{tab:front3}
\end{tabular}
\end{table}

\section{Conclusion}
Highly correlated relativistic coupled cluster theory has been employed to study the oscillator strengths 
of the astrophysically 
important electric dipole transitions. To our knowledge these are the first calculations of oscillator 
strengths for most of the transitions presented here. All the transitions are in the ultraviolet, visible 
or near infrared regions. The good agreement between transition amplitudes obtained from length and velocity
gauge expressions highlight the accuracies of our computational method. This work will 
motivate astronomers to observe these lines of V$^{4+}$ to predict the abundances of these species in 
astronomical bodies and experimentalists to verify our results.

\section{Acknowledgment}
We would like to acknowledge Prof. B. P. Das and Prof. Debashis Mukherjee for helpful discussions.
The work reported in this paper has been supported by the `New Faculty Scheme' project by IC\&SR, 
IIT-Madras. The computations were carried out by the computer financed by this 
project.

\end{document}